\input ./style/arxiv-general.cfg
\documentclass[aoas,MSNbibl,nameyear,seceqn,dvips]{arximspdf}
\makeatletter
   \@ifpackageloaded{graphicx}{}{\usepackage{graphicx}}
\makeatother
\usepackage[linesnumbered,boxruled]{algorithm2e}
\usepackage{url,breakurl}

\doi{10.1214/14-AOAS758}
\volume{9}
\issue{1}
\pubyear{2015}
\firstpage{27}
\lastpage{42}
\docsubty{FLA}

\makeatletter
 \let\my@algocf@latexcaption\algocf@latexcaption
\let\my@addcontentsline\addcontentsline
\long\def\algocf@latexcaption#1[#2]#3{%
\def\addcontentsline##1##2##3{}%
\my@algocf@latexcaption{#1}[#2]{#3}%
\global\let\addcontentsline\my@addcontentsline%
}
\newcommand{\rrVert}{\Vert}
\newcommand{\llVert}{\Vert}

\newcommand{\eqref}[1]{(\ref{#1})}

\newtheorem{theorem}{Theorem}
\makeatother

\begin{document}
\begin{frontmatter}

\title{Convex hierarchical testing of interactions}
\runtitle{Convex hierarchical testing of interactions}

\begin{aug}
\author[A]{\fnms{Jacob}~\snm{Bien}\ead[label=e1]{jbien@cornell.edu}},
\author[B]{\fnms{Noah}~\snm{Simon}\ead[label=e2]{nrsimon@uw.edu}}
\and
\author[C]{\fnms{Robert}~\snm{Tibshirani}\corref{}\thanksref{T1}\ead[label=e3]{tibs@stanford.edu}}
\runauthor{J. Bien, N. Simon and R. Tibshirani}
\affiliation{Cornell University, University of Washington and Stanford
University}
\address[A]{J.~Bien\\
Department of Biological Statistics\\
\quad and Computational
Biology\\
and\\
Department of Statistical Science\\
Cornell University\\
Ithaca, New York 14853\\
USA\\
\printead{e1}}
\address[B]{N. Simon\\
Department of Biostatistics\\
University of Washington\\
Seattle, Washington 98195\\
USA\\
\printead{e2}}
\address[C]{R. Tibshirani\\
Department of Health  Research and Policy\\
and\\
Department of Statistics\\
Stanford University\\
Stanford, California 94305\\
USA\\
\printead{e3}}
\end{aug}
\thankstext{T1}{Supported in part by NSF Grant
DMS-9971405 and
National Institutes of Health Contract N01-HV-28183.}

%
\received{\smonth{7} \syear{2013}}
%
\revised{\smonth{4} \syear{2014}}

%
\begin{abstract}
We consider the testing of all pairwise interactions in a two-class
problem with many features.
We devise a hierarchical testing framework that considers
an interaction only when one or more of its constituent features has a
nonzero main effect.
The test is based on a convex optimization framework that seamlessly
considers main effects and interactions together.
We show---both in simulation and on a genomic data set from the
SAPPHIRe study---a~potential gain in power and interpretability
over a standard (nonhierarchical) interaction test.
\end{abstract}

%
\begin{keyword}
\kwd{Interactions}
\kwd{testing}
\kwd{lasso}
\end{keyword}
\end{frontmatter}

\section{Introduction}\label{sec1}

We consider the standard two-class problem with $y_i\in\{1,2\}$ and
$p$ features
$\{x_{i1}, x_{i2}, \ldots, x_{ip}\}$ measured on each of $i=1,2,\ldots,
n$ observations.
Large-scale hypothesis testing for the effects of individual features
(such as genetic markers; see Section~\ref{sec5})
is a challenging problem and has received much attention in recent
years [e.g., \citet{Efron2010,DV2008}].
The problem of testing for interactions between pairs of features is even
more difficult, as there are $p\choose2$ interactions.
\citet{BLT2011} show that standard
permutation tests cannot be used for interaction testing (because the
correct null hypothesis is difficult to enforce) and
propose
instead a parametric bootstrap-based approach. \citet{ST2012}
devise
a permutation approach that exploits the close relationship between the
``forward'' logistic model (based on $Y|X$) and a ``backward''
discriminant analysis (Gaussian)
model (based on $X|Y$).

When $p$ is large, the large number of potential pairwise interactions can
result in low power for detecting the true effects. One strategy used
by data analysts is to first screen the data for significant main
effects, and then to  test for interactions only among those features that
are themselves significant. This approach can be effective, but it has
some drawbacks. Specifically, at what threshold does one stop entering
main effects? And should this
threshold vary across main effects depending
on
the strength
of the interactions?

The above two-stage strategy can be viewed as ``hierarchical'':
Interactions are considered only if both constituent
main effects are significant.
In this paper we propose a convex formulation that models main effects
and interactions together, in a hierarchical fashion. It provides a
testing framework that
seamlessly combines main effects and interactions. We call the method
\textit{convex hierarchical testing} (CHT).
The method is closely related to the recently proposed hierarchical
lasso regression method
(``hierNet'') of \citet{BTT2013}. A difference is that CHT seeks
marginal interactions while hierNet looks for conditional
interactions. We focus exclusively on pairwise
interactions in the paper  but discuss possible extensions to higher
order interactions in Section~\ref{sec:discussion}.

\section{Testing interactions using a convex formulation}\label{sec2}
Our proposal has three main components. The first is
to define interactions and main effects in terms of a ``backward
model'' [\citet{ST2012}]. The second component is to relate the testing
problem to the solution path of a convex optimization
problem. The third component is to introduce hierarchical sparsity
within the convex optimization framework. We present these
components in the next three subsections.

\subsection{Defining interaction and main effects via the backward model}
\label{sec:backward}

A common approach to testing interactions is to consider ${p\choose
2}$ logistic regression models of the form
\[
\log \biggl[\frac{p(Y|X_j,X_k)}{1-p(Y|X_j,X_k)} \biggr]=\alpha _0+
\alpha_jX_j+\alpha_kX_k+
\alpha_{jk}X_jX_k
\]
and then to test whether $\alpha_{jk}=0$. However, \citet{ST2012} argue
that this definition of interaction is less natural than one coming
from considering a
``backward model'' in which the feature vector $X\in\mathbb{R}^p$ is
modeled conditional on the class label $Y\in\{1,2\}$:
\[
X|Y=\ell\sim N_p \bigl(\mu^{(\ell)},\Sigma^{(\ell)}
\bigr).
\]
In particular, they redefine an interaction between $X_j$ and $X_k$ to
mean that $\operatorname{Cor}(X_j,X_k|Y=\ell)=\rho_{jk}^{(\ell)}=(\Sigma
_{jj}^{(\ell)}
\Sigma_{kk}^{(\ell)})^{-1/2}\Sigma_{jk}^{(\ell)}$ depends on $\ell$. Their
main criticism of defining interactions based on the forward model is
that if the \textit{marginal} quantity $\operatorname{Var}(X_j|Y=\ell)=\Sigma
_{jj}^{(\ell)}$ depends on $\ell$,
then $\alpha_{jk}\neq0$ for all $k$. This does not correspond to what
a biologist, say, would consider an interesting interaction because it
is not a property of the \textit{pair} of variables $(X_j,X_k)$. Likewise,
a natural definition for main effects in the backward model
is a difference in class mean for that variable. Hence, we work in
the backward model and test hypotheses of two kinds, which we will
refer to as ``main effects'' and ``interactions'':
\begin{eqnarray*}
&& H_{0,j}\dvtx \mu_{j}^{(1)} = \mu_{j}^{(2)}
\qquad\mbox{for }1\le j\le p,
\\
&& H_{0,jk}\dvtx \rho_{jk}^{(1)} = \rho_{jk}^{(2)}
\qquad\mbox{for $1\le j<k\le p$}.
\end{eqnarray*}
For testing $H_{0,j}$, a common choice would be the standard $t$-statistic,
\[
w_j = \frac{\bar{x}_{j}^{(1)} - \bar{x}_{j}^{(2)}}{\sqrt
{s_j^{(1)2}/n_1 + s_j^{(2)2}/n_2}},
\]
where\vspace*{1pt} $\bar x_j^{(\ell)}=n_\ell^{-1}\sum_{i\dvtx y_i=\ell}x_{ij}$ and
$s_j^{(\ell)2}=(n_\ell-1)^{-1}\sum_{i\dvtx y_i=\ell}(x_{ij}-\bar
x_j^{(\ell)})^2$
are the sample means and variances within class $\ell$. For testing
$H_{0,jk}$, a common choice would be based on the difference of the Fisher
transformed sample correlations between the two classes:
\[
z_{jk}= \biggl(\frac{1}{n_1-3}+\frac{1}{n_2-3}
\biggr)^{-1/2} \bigl[\operatorname{arctanh}\bigl(\hat\rho_{jk}^{(1)}
\bigr)-\operatorname{arctanh}\bigl(\hat\rho _{jk}^{(2)}\bigr) \bigr].
\]
Here,\vspace*{1pt}
$\hat\rho_{jk}^{(\ell)}=(n_\ell-1)^{-1}\sum_{i\dvtx y_i=\ell
}(x_{ij}^{(\ell)}-\bar x_j^{(\ell)})(
x_{ik}^{(\ell)}-\bar x_k^{(\ell)})/(s_j^{(\ell)} s_k^{(\ell)})$ is
the sample correlation within class
$\ell$. Both $w_j$ and $z_{jk}$ are scaled so that they are approximately
standard normal (for large $n_1$ and $n_2$).

\subsection{Test statistics through convex optimization}
\label{sec:cvx-test}

We would like to select interactions based on the size of $|z_{jk}|$ but
also somehow give a ``boost'' to interactions whose main effects are
large.
One could try to achieve this through a two-stage procedure in which one
first screens the individual features and then tests for interactions
only among those features selected at the first phase. This kind of
method is explored, for example, in \citet{GEPI:GEPI20300},
\citet{GEPI:GEPI21610} and \citet{WZ2009}. However, such an approach
to the
hierarchy requirement can lead to algorithmic shortsightedness. In
particular, a very strong interaction
will be ignored if the corresponding main effects fail to make the
threshold in the
first phase. We seek a method that enforces the hierarchy constraint
but \textit{jointly} considers which interactions and main effects to
include in the
model.

Suppose\vspace*{1pt} that we define a testing procedure through a convex
optimization problem involving both $w_{j}$ and $z_{jk}$.
Let $\beta^+,\beta^-\in\mathbb{R}^p$ and $\theta\in\mathbb
{R}^{p(p-1)}$ be
optimization variables. Given the objective function
%
\begin{eqnarray}
\label{eq:allpairs}
L_{\lambda}\bigl(\beta^+,\beta^-,\theta\bigr)&=&
\frac{1}{2}\sum_{j=1}^p
\bigl(w_j-\bigl(\beta^+_j-\beta^-_j\bigr)
\bigr)^2+\frac{1}{2}\sum_{j=1}^p
\sum_{k\dvtx k\neq
j}(z_{jk}-\theta_{jk})^2
\nonumber
\\[-8pt]
\\[-8pt]
\nonumber
&&{}+\lambda\sum_{j=1}^p\bigl[
\beta^+_j +\beta^-_j\bigr]+ \lambda\sum
_{j=1}^p\sum_{k\dvtx k\neq j}|
\theta_{jk}|,
\end{eqnarray}
we may define the problem
\[
\min_{\beta^+,\beta^-,\theta}L_{\lambda}\bigl(\beta^+,\beta^-,\theta \bigr)
\quad \mbox{s.t.}\quad \beta^+_j\ge0,\beta^-_j\ge0 \mbox{ for }1
\le j\le p,
\]
where $\lambda$ is a tuning parameter. For each fixed $\lambda\ge0$,
the pair
$(\hat{\beta}^+(\lambda)-\hat{\beta}^-(\lambda),\hat\theta
(\lambda))$ is unique.
Consider the path of solutions formed by varying $\lambda$ from
$\infty$ to $0$. The solution path goes from
$(0,0)\in\mathbb{R}^{p+p(p-1)}$ to
$(w,z)$ and is piecewise-linear with knots at the values of $\lambda$
for which individual coordinates
of $\hat\theta_{jk}(\lambda)$ or $\hat{\beta}^+_j(\lambda)-\hat
{\beta}^-_j(\lambda)$
become nonzero. It is straightforward\vadjust{\goodbreak} to show that these knots occur
precisely at the values of the standard test statistics introduced in the
previous section:
%
\begin{equation}\label{eqn:knots}
|w_1|,\ldots,|w_p|,\qquad |z_{12}|,
\ldots,|z_{p-1,p}|.
\end{equation}
This observation suggests how a regularized regression problem can be
viewed as producing test statistics: One defines the test statistic
associated with a variable to be the $\lambda$ value at which the
corresponding parameter becomes nonzero.

Now in this setup, the $k$th knot is just equal to the $k$th largest value
among those in (\ref{eqn:knots}), so our test for interactions is just
the usual one,
based on the size of $|z_{jk}|$. We have not obtained anything new.
To exploit hierarchy, we will modify the optimization problem as
described in the next section.

\subsection{Convex hierarchical testing}
\label{sec:conv-hier-test}
The procedure described above does not share information
between main effects and interactions.
Our proposal in this paper is to add a convex hierarchy constraint to the
problem, which will lead to \textit{main-effect ``informed'' thresholds}
for testing the interactions (and likewise interaction ``informed'' thresholds
for testing main effects).

\citet{BTT2013} develop a hierarchical interactions lasso method
in the forward model. The hierarchical sparsity is achieved by adding
a set of convex
constraints to the lasso problem. We may similarly impose hierarchy
in the backward model by modifying \eqref{eq:allpairs} to get a
hierarchical interactions test in the
backward model:
%
\begin{eqnarray}
\bigl(\hat{\beta}^+,\hat{\beta}^-,\hat{\theta}\bigr)&=& \arg\min
L_{\lambda
}\bigl(\beta^+,\beta^-,\theta\bigr)\quad \mbox{subject to}\quad
\beta^+_j,\beta^-_j\ge0,
\nonumber
\\[-8pt]
\label{eq:hier}
\\[-8pt]
\nonumber \sum
_{k\dvtx k\neq j}|\theta _{jk}| &\leq& \beta^+_j+
\beta^-_j.
\end{eqnarray}
We solve this problem for all $\lambda$ and define the test statistic
associated with an interaction to be the $\lambda$ value at which the
corresponding parameter becomes nonzero.
This is the main proposal of this paper.

The addition of the constraint imposes a ``budget'' $\beta_j^+ + \beta
_j^-$ on
the total interactions that involve feature $j$.
In particular, the constraint $\sum_{k\dvtx k\neq j}|\theta_{jk}|\leq
\beta^+_{j}+\beta^-_{j}$
implies that at least one of $\beta^+_{j}$ and $\beta^-_{j}$ must be
nonzero in order for $\theta_{jk}$ to be nonzero. Although in
theory we could have $\hat\beta^+_{j}=\hat\beta^-_{j}$ with both
values positive, this
happens with probability zero under reasonable assumptions [\citet
{BTT2013}]. As a result,
$\hat\theta_{jk}\neq0$ implies $\hat\beta_j \neq0$, and similarly for
$\hat\theta_{kj}$. Thus, the $jk$th interaction is nonzero if at
least one of $\hat\beta_j$ or $\hat\beta_k$ is nonzero.
This property has been called \textit{weak hierarchy} [see, e.g., \citet{BTT2013}], in
contrast to \textit{strong hierarchy}, which requires both $\hat\beta_j$ and
$\hat\beta_k$ to be nonzero
in order for $\hat\theta_{jk}$ to be nonzero. Problem (\ref{eq:hier}) is convex, due to the fact that we have represented each
main effect $\beta_j$ as the difference of
two nonnegative quantities $\beta_j^+, \beta_j^-$. It would not
be convex if we had used $|\beta_j|$ in place of $\beta_j^+ + \beta
_j^-$ in the
constraint above. This is because the set $\{(x,t)\dvtx \|x\|_1\le|t|\}$ is
not convex.

Working in the optimization-based testing framework of the previous
section, we consider the solution path
(parameterized by $\lambda$) of this problem and then define
the test statistics for interactions and main effects to be the
$\lambda$ values at which these values become nonzero (i.e., the knots
of the path). In\vspace*{1pt}
particular, for testing the $jk$th interaction, we take the largest
$\lambda$ for which either $\hat\theta_{jk}$ or
$\hat\theta_{kj}$ is nonzero, and for testing the $j$th main effect
we compute
the largest $\lambda$
for which either
$\hat{\beta}^+ - \hat{\beta}^-$
is
nonzero. That is, letting $\hat\beta(\lambda)=\hat{\beta
}^+(\lambda)-\hat{\beta}^-(\lambda)$ and
$\hat\theta(\lambda)$ denote the solution as a function of $\lambda
$, our proposed test statistics are
\begin{eqnarray*}
\hat{\lambda}_{j}&=&\sup\bigl\{\lambda\ge0\dvtx \hat{\beta}_{j}(
\lambda )\neq0\bigr\},
\\
\hat{\lambda}'_{jk}&=& \max\{\hat\lambda_{jk},
\hat\lambda_{kj}\},
\end{eqnarray*}
where
%
\begin{equation}\label{eqn:teststat}
\hat{\lambda}_{jk}=\sup\bigl\{\lambda\ge0\dvtx \hat{\theta}_{jk}(
\lambda )\neq 0\bigr\}.
\end{equation}

In Lemma~2 of the online supplement [\citet{Bien14supp}], we prove that
\eqref{eq:hier} has a unique solution
for each $\lambda>0$, so $\hat\lambda_j$ and $\hat\lambda_{jk}$ are
well defined. Furthermore, we prove in Proposition 2
of the online supplement [\citet{Bien14supp}] that
$|\hat\theta_{jk}(\lambda)|$ is nonincreasing in $\lambda$, which means
that $\hat\lambda_{jk}$ is in fact the unique point in the path where
$\hat\theta_{jk}(\lambda)$ becomes nonzero.

Without the hierarchy constraints in \eqref{eq:hier}, we would have
$\hat\lambda_{jk}=|z_{jk}|$ and $\hat\lambda_j=|w_j|$ as in Section~\ref{sec:cvx-test}.
The weak hierarchy property of the solution to \eqref{eq:hier} implies that
\[
\hat{\lambda}'_{jk}\le\max\{\hat{\lambda}_j,
\hat{\lambda}_k\}.
\]
While one might assume that finding the knots of
\eqref{eq:hier} would be computationally intensive, requiring one to
solve the problem at many
values of $\lambda$, it turns out that there is an explicit
analytical form for the knots of this path, meaning that computing
the test statistics is in fact computationally simple.

\begin{theorem}\label{thm1}
The knots of the solution path of \eqref{eq:hier} have the following
closed-form expressions:
%
\begin{eqnarray}
\hat\lambda_j&=& \max \biggl\{|w_j|,\frac{|w_j|+\Vert z_{j\cdot
}\Vert_\infty}{2}
\biggr\},
\nonumber
\\[-8pt]
\label{eq:teststat}
\\[-8pt]
\nonumber
\hat\lambda_{jk}&=& \min \biggl\{|z_{jk}|,
\frac{|z_{jk}|}{2}+\frac{ [|w_j|-\llVert \mathcal S
(z_{j\cdot},|z_{jk}| )\rrVert _1 ]_+}{2} \biggr\},
\end{eqnarray}
where $z_{j\cdot}=\{z_{jk}\dvtx k\neq j\}\in\mathbb{R}^{p-1}$ is the
vector of
interaction contrasts involving the $j$th variable and $\mathcal S$ is the
soft-thresholding function so that $\|\mathcal S(z_{j\cdot},\break |z_{jk}|)\|
_1=\sum_{\ell\dvtx |z_{j\ell}|>|z_{jk}|} (|z_{j\ell
}|-|z_{jk}| )$.
\end{theorem}
\begin{pf}
See Proposition 4 in the online supplement [\citet{Bien14supp}].
\end{pf}
These formulae are somewhat complex, but we can interpret them loosely
as follows.
Each main effect is ``boosted'' by the size of the largest interaction
in its row  due to the hierarchy constraint. In contrast, each
interaction is shrunk by as much as half of its size, with the
shrinkage amount
less when the main effect is large or the interaction is large relative to
the other interactions in that row. Interestingly, $\hat\lambda_{jk}$
depends only on $w_j$ and on  those interactions in the $j$th row that are
at least as large (in absolute value) as $z_{jk}$.

At one extreme, suppose $w_j=0$. In this case,
$\hat\lambda_j=\|z_{j\cdot}\|_\infty/2$ and
$\hat\lambda_{jk}=|z_{jk}|/2$ (compared to the nonhierarchical
statistics, which would be 0 and $|z_{jk}|$).
On the other extreme, $|w_j|\gg\|z_{j\cdot}\|_1$, in which case
$\hat\lambda_j=|w_j|$ and $\hat\lambda_{jk}=|z_{jk}|$ (which are
identical to the nonhierarchical statistics).

\begin{figure}

\includegraphics{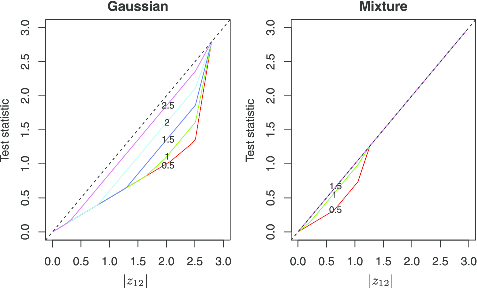}

\caption{Graphical illustration of formula
\protect\eqref{eq:teststat} for two different distributions of interactions
(two panels) and different size of main effects $w$ (colored lines).
Broken line is the $45^{\circ}$ line. Figures show how test statistic
$\hat\lambda_{12}$ varies with $|z_{12}|$.
Full details in text.}
\label{fig:intplot}
\end{figure}

Figure~\ref{fig:intplot} gives a graphical illustration of the formula
in \eqref{eq:teststat}.
We set the number of interactions to $50$.
In the left panel the interaction contrasts $z_{1k}$, for $k>2$, are
generated as $N(0,1)$.
The plot shows the test statistic $\hat\lambda_{12}$ as a function of
$|z_{12}|$
and the main effect $w_1$ (different colored curves with main effect
indicated), along with the $45^\circ$ line.
We see that the interaction effect is shrunk substantially until it reaches
about 2.75  and that the amount of shrinkage is less when the main effect is larger.
In the right panel there are $49$ small interactions distributed as
$N(0,0.5^2)$ and one large interaction whose value varies along the
horizontal axis. Now we see that there is shrinkage
only until a value of about 1.5
and that
a main
effect of 1.5 is sufficient to ensure no shrinkage at all.

The knot-based test statistics produce a single ranking of all interactions
and main effects. Our test rejects any null hypotheses whose
corresponding knots are greater
than a threshold. This threshold is chosen to meet a
desired false discovery rate (FDR). In Section~\ref{sec:fdr}, we give
a method for
estimating the FDR. In this way, a practitioner can
choose a cutoff with an acceptable FDR. As mentioned above, we call our
method \textit{convex hierarchical testing} (CHT). Algorithm \ref{alg:CHT} spells out
the full
procedure, which consists of computing the test statistics and then
estimating the FDR at a series of cutoffs. The corresponding version
of this proposal given in~\eqref{eq:allpairs} that does not have the hierarchy
constraints we call the \textit{all-pairs} method.

\begin{algorithm}[b]
\caption{Algorithm for convex hierarchical testing}
\label{alg:CHT}
\KwIn{Main effect and interaction contrasts,
$w_1,\ldots,w_p$ and $z_{jk}$ for $1\le j,k\le p$, $j\neq k$, as
defined in
Section~\ref{sec:backward} and a threshold $\bar\lambda$.} Compute
$\hat\lambda_j$ for $1\le
j\le p$ and $\hat\lambda'_{jk}$ for $1\le j<k\le p$ using
\eqref{eq:teststat}.


Reject all hypotheses $H_{0,jk}$ for
which $\hat\lambda'_{jk}\le\bar\lambda$ (and, if main effects are
of interest, all $H_{0,j}$ for which
$\hat\lambda_j\le\bar\lambda$).

Repeat $B$ times: do steps 1--2 on data permuted as described in Section~\ref{sec:fdr}.

Use \eqref{eq:fdr-hat} to compute $\widehat{\mathrm{FDR}}(\bar\lambda)$.
\end{algorithm}

\section{Some insight into the optimization problem 
\texorpdfstring{\protect\eqref{eq:hier}}{(2.3)}}
\label{sec:details}
Although the ranking of interactions from the above procedure comes
from a seemingly complicated
optimization problem,
the solutions actually have a simple form.
In particular,
we prove in the online supplement [see Lemma~1 of \citet{Bien14supp}] that
%
\begin{eqnarray}
\label{eqn:soft} \hat\theta_{jk}(\lambda)&=&\mathcal S
\bigl(z_{jk},\lambda+\hat\alpha_j (\lambda)\bigr),
\nonumber
\\[-8pt]
\\[-8pt]
\nonumber
\hat{\beta}^+_j (\lambda)-\hat{\beta}^-_j (\lambda)&=&
\mathcal S\bigl(w_j,~\lambda-\hat\alpha_j(\lambda)
\bigr).
\end{eqnarray}
Here $\mathcal S(x,t)=\operatorname{sign}(x) \cdot (|x|-t)_+$ is the
soft-thresholding function,
and the value $\hat\alpha_j(\lambda)\in[0,\lambda]$ emerges from
the solution
to problem \eqref{eq:hier},
with $\hat\alpha_j(\lambda)=0$ if the hierarchy constraint
$\sum_{k\dvtx k\neq j}|\hat\theta_{jk}|\leq\hat{\beta}^+_{j}+\hat
{\beta}^-_{j}$
is loose (i.e., a strict inequality).

For the all-pairs problem following \eqref{eq:allpairs}, $\hat\alpha
(\lambda)=0$ gives
the solution. Thus, we can think of $\hat\alpha_j(\lambda)$ as the bridge
between the main effects and interactions that ensures hierarchy. Its
value depends on both the interactions and the main effects. It is
easy to see from \eqref{eqn:soft} that the $j$th main effect becomes
nonzero at the knot
$\hat\lambda_j=|w_j|+\hat\alpha_j(\hat\lambda_j)$ and the $jk$th
interaction becomes
nonzero at $\hat\lambda_{jk}=|z_{jk}|-\hat\alpha_j(\hat\lambda_{jk})$.
Thus, the solution path $\hat\alpha_j(\lambda)\ge0$ is responsible for
the hierarchy-related ``boost'' we described in the
introductory section.

When $|w_j|$ is large enough relative to the $|z_{jk}|$'s,
$\hat\alpha_j(\lambda)=0$, that is, hierarchy holds automatically. When
$|z_{jk}|$ is large relative to $|w_j|$, then we may have
\mbox{$\hat\alpha_j(\lambda)>0$}, and this can be as large as $\lambda$.
From \eqref{eqn:soft}, we see that $\hat\alpha_j(\lambda)>0$ means
that $|w_j|$ are shrunk by less [or even\vadjust{\goodbreak} not at all if
$\hat\alpha_j(\lambda)=\lambda$] and that the interactions are
shrunk by more (up to twice as much as in the all-pairs approach).
This gives some intuition for Theorem~\ref{thm1}.

\section{A simulation study}\label{sec4}
We simulate Gaussian data from the backward model with $n=200$
observations and $p=50$ features in
two classes $y\in\{1,2\}$. In all cases, we take $\mu^{(1)}=0$ and
$\Sigma^{(1)}=I_p$. We consider six scenarios, each of which has 10
nonzero interactions:
\begin{itemize}
\item \textit{Weak Hierarchical Truth (small interactions)}: We take
$\mu^{(2)}_j=2$ for $j=1,\ldots,5$, and then
select a random 10 interactions $(j_i,k_i)\in\{1,\ldots,5\}\times\{
6,\dots,p\}$ to be nonzero:
\[
\Sigma^{(2)}_{jk}= %
\cases{ 0.3, &\quad$\mbox{if
}(j,k)=(j_i,k_i) \mbox{ for some }i=1,\ldots,10$,
\vspace*{3pt}
\cr
\Sigma^{(1)}_{jk}, &\quad$\mbox{otherwise}$.}
\]
\item \textit{Weak Hierarchical Truth}: Same as above, but with
0.5 instead of 0.3.
\item \textit{Strong Hierarchical Truth}: We take
$\mu^{(2)}_j=2$ for $j=1,\ldots,5$, and then take
\[
\Sigma^{(2)}_{jk}=
\cases{0.5, &\quad$\mbox{if }1
\le j,k\le5,j\neq k$,\vspace*{3pt}
\cr
\Sigma^{(1)}_{jk}, & \quad\mbox{otherwise}.}
\]

\item \textit{No Main Effects Truth}:  Same as Strong Hierarchical Truth
except $\mu^{(2)}=0$.
\item \textit{No Main Effects Truth (large interactions)}: Same as above,
but with 0.9 instead of 0.5.
\item \textit{Anti-Hierarchical Truth}: We take
$\mu^{(2)}_j=2$ for $j=1,\ldots,5$, and then take
\[
\Sigma^{(2)}_{jk}= %
\cases{ 0.5, & \quad$\mbox{if }6
\le j,k\le10,j\neq k$,\vspace*{3pt}
\cr
\Sigma^{(1)}_{jk},& $\quad\mbox{otherwise}$.}
\]
\end{itemize}
We compare CHT with the all-pairs testing
procedure, along with two different two-stage screening methods: In the
``strong'' version we retain all main effects
with $z$ scores above the 75th percentile  and then in the second stage
test for
interactions only among the retained variables.
In the weak version, we consider all interactions among pairs of variables
where at least one variable has a $z$ score above the 75th percentile.

\begin{figure}[t]

\includegraphics{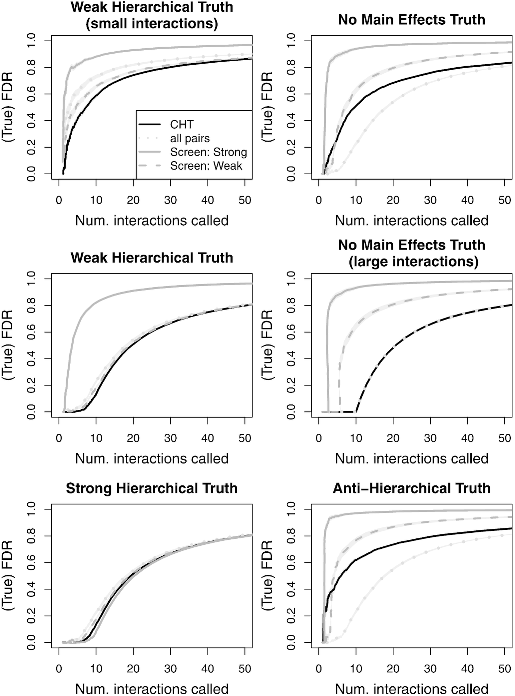}

\caption{(True) false discovery rates of four different
procedures over six different settings. Error bars are in light
gray and mostly too narrow to be seen. CHT and the screening methods
do better than all-pairs
when the truth is hierarchical (left column). When there are no main
effects but there are large interactions
(middle right), CHT and all-pairs are able to perfectly identify all
interactions, whereas the screening methods do not.}
\label{fig:testing}
\end{figure}

\begin{figure}[t]

\includegraphics{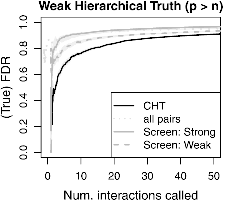}

\caption{(True) false discovery rates of four different
procedures when the truth is weak hierarchical with $p=100$, $n=50$.
Only ten
of the 4950 interactions are actually nonzero.}
\label{fig:pgn}
\end{figure}

Figure~\ref{fig:testing} 
shows the true (as opposed to the estimated) false discovery
rate for testing interactions, averaged over 100 simulations.
%
In the weak hierarchical scenario with small interactions, we see that CHT
shows a substantial improvement in FDR over all-pairs, with the weak
screen method performing
a little worse. In the weak hierarchical scenario with larger
interactions, the same ordering of methods holds, although the
differences are less pronounced. In the strong hierarchical truth
scenario, the strong screening rule does best (by a small amount). We
see that in all other scenarios, the strong screening rule does
unacceptably poorly. In the three scenarios where hierarchy does not
hold, all-pairs does best. When no main effects are present and the
interactions are large (middle right), CHT does as well as all-pairs.
This behavior can be explained by \eqref{eq:teststat}: When all main effects
are small enough, we have $\hat\lambda_{jk}\approx|z_{jk}|/2$, which
has the same ordering as all-pairs. For the screening methods, on the
other hand, if a main effect is small, large interactions can go
completely undetected. In the anti-hierarchical setting, we construct a
scenario in which the
hierarchy assumption is explicitly violated. Not surprisingly, CHT and the
screening methods do poorly compared to all-pairs. Figure~\ref{fig:pgn}
shows the performance of the methods in a scenario
identical to the ``Weak Hierarchical Truth'' but with $p=100$ and
$n=50$. There are still only ten nonzero interactions, but now there
are 4950 interactions to choose among. The high FDR
values show that this is a more challenging scenario; however, CHT
performs well compared to the other methods.

\begin{figure}

\includegraphics{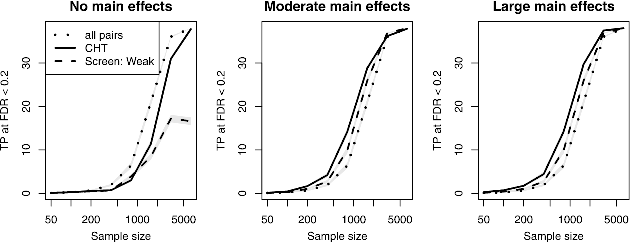}

\caption{Average number of true positives (i.e., nonnull
interactions called significant)
with FDR${}<{}$0.2 (over 50 replications), for varying sample sizes
(horizontal axis, logarithmic scale) and size of the main effect. True
number of nonnull interactions is 40.}
\label{fig:power}
\end{figure}

In Figure~\ref{fig:power}, we vary the strength of the main effects in
a weak hierarchical scenario with 40 nonzero interactions. We compare
all-pairs, CHT and weak-screen in their
ability to correctly detect interactions while controlling FDR at a
given sample size. We estimate the average
number of nonnull interactions called significant (over 50 replications)
with FDR${}<0.2$, for varying sample sizes (horizontal axis) and size of
the main effect (varying across panels).
When no main effects are present, the all-pairs method does best, and
CHT does much better than weak-screen (which is unable to detect over
half of the interactions regardless of increasing sample size because
these interactions have main effects that are too small).
In the other two scenarios, CHT does best.

\section{Real data example: SAPPHIRe study data}\label{sec5}
This data set was analyzed in \citet{Park01012008}, following the
study of
\citet{Huang2004}.
The study sought to find genes associated with hypertension.
A sample of 580 Chinese women, 216 hypotensive and
364 hypertensive, were studied. The predictors (see Table~\ref{tab:sapphire1}) are menopausal and insulin resistance statuses as
well as genotypes on 21 distinct loci.
\begin{table}[b]
\caption{List of predictors in the SAPPHIRe data set}
\label{tab:sapphire1}
\begin{tabular*}{\tablewidth}{@{\extracolsep{\fill}}lccc@{}}\hline
\textbf{Predictor} \textbf{number} & \textbf{Name} & \textbf{Predictor} \textbf{number} & \textbf{Name}\\
\hline
\phantom{0}1& \texttt{Reached menopause? }&14& \texttt{PTPN1i4INV} \\
\phantom{0}2& \texttt{insulin t=-10 } & 15& \texttt{Cyp11B2x1INV} \\
\phantom{0}3 & \texttt{insulin t=60} & 16& \texttt{PTPN1x9INV} \\
\phantom{0}4 & \texttt{insulin t=120 }& 17& \texttt{ADRB3W1R} \\
\phantom{0}5 & \texttt{HUT2SNP5} & 18&\texttt{KLKQ3E} \\
\phantom{0}6 & \texttt{HUT2SNP7} &19 & \texttt{AGT2R1A1166C} \\
\phantom{0}7 & \texttt{BADG16R} &20& \texttt{AVPR2G12E} \\
\phantom{0}8 & \texttt{AVPR2A1629G} &21& \texttt{MLRI2V} \\
\phantom{0}9 & \texttt{AGT2R2C1333T} &22& \texttt{AGTG6A} \\
10& \texttt{PPARG12 } &23& \texttt{Cyp11B2-5paINV} \\
11& \texttt{CD36x2aINV } &24& \texttt{PTPN1i1} \\
12& \texttt{MLRi6INV } & 25& \texttt{PTPN1i4} \\
13& \texttt{Cyp11B2i4INV} \\
\hline
\end{tabular*}
\end{table}

\begin{figure}

\includegraphics{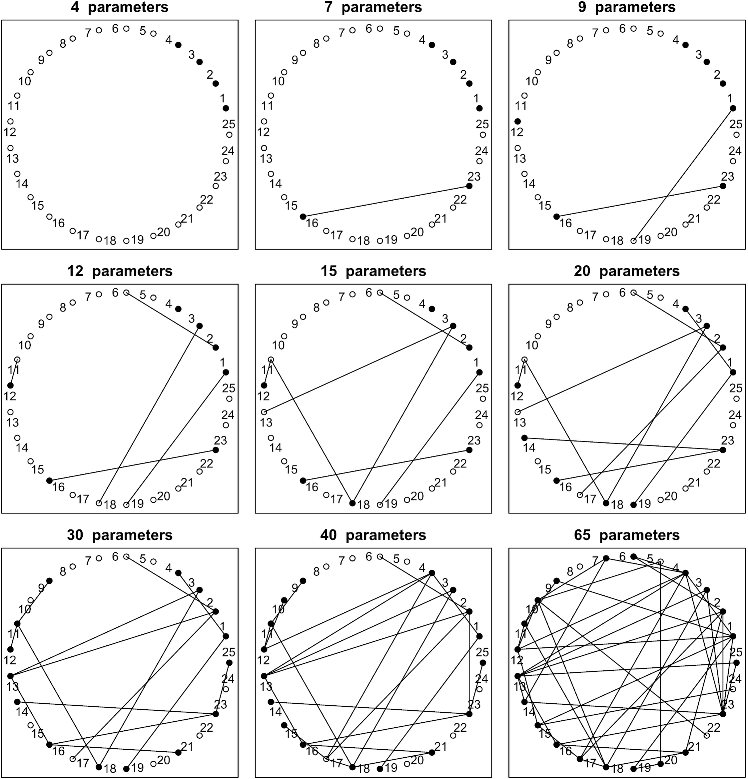}

\caption{Convex hierarchical testing: main effects (black dots)
and interactions (edges)  for 9 different decreasing values of
$\lambda$. Weak hierarchy ensures that each edge is incident to at
least one black dot.}
\label{fig:wheelplot}
\end{figure}

The first four predictors (all nongenetic) have the strongest effects
individually, although
none were
(\url{http://www.grammarmudge.cityslide.com/articles/article/1026513/9903.htm})
 significantly different across the
two groups (details not shown).
Table~\ref{tab:sapphire2}
shows the first ten interactions found by the all-pairs and
CHT methods.
%
\begin{table}
\caption{Top ten interactions found by all-pairs and
convex hierarchical test methods}
\label{tab:sapphire2}
\begin{tabular*}{\tablewidth}{@{\extracolsep{\fill}}lc@{}}\hline
\textbf{All-pairs}&\textbf{Convex hierarchical testing}\\
\hline
 \texttt{PTPN1x9INV:Cyp11B2-5paINV} & \texttt{PTPN1x9INV:Cyp11B2-5paINV}\\
 \texttt{CD36x2aINV:MLRi6INV} & \texttt{Reached menopause?:AGT2R1A1166C}\\
 \texttt{Cyp11B2-5paINV:PTPN1i4} & \texttt{CD36x2aINV:MLRi6INV}\\
 \texttt{PTPN1i4INV:Cyp11B2-5paINV} & \texttt{insulin t=60:KLKQ3E}\\
 \texttt{Cyp11B2i4INV:PTPN1x9INV} & \texttt{insulin t=-10:HUT2SNP7}\\
 \texttt{CD36x2aINV:KLKQ3E} & \texttt{CD36x2aINV:KLKQ3E}\\
 \texttt{PTPN1x9INV:MLRI2V} & \texttt{insulin t=60:Cyp11B2i4INV}\\
 \texttt{Reached menopause?:AGT2R1A1166C} & \texttt{insulin t=-10:ADRB3W1R}\\
 \texttt{Cyp11B2i4INV:PTPN1i4} & \texttt{PTPN1i4INV:Cyp11B2-5paINV}\\
 \texttt{AGT2R2C1333T:CD36x2aINV} & \texttt{Reached menopause?:insulin t=120}\\
\hline
\end{tabular*}
%
\end{table}
Five interactions are shared across these lists. It is interesting to
observe how these lists are similar and how they are different. Every
gene--gene interaction found by CHT is also in the all-pairs list. Every
interaction found \textit{only} by all-pairs and not by CHT is a gene--gene
interaction, while every interaction found by CHT but not by all-pairs
involves at least one nongenotype predictor.
Figure~\ref{fig:wheelplot} 
depicts the main effects and interactions found by CHT
for different values of the regularization parameter $\lambda$.
\begin{table}
\caption{Ten most frequent interactions found by all-pairs,
weak-screening and CHT methods over 100~bootstrap replications}
\label{tab:sapphire3}
\begin{tabular*}{\tablewidth}{@{\extracolsep{\fill}}lc@{}}
\hline
\textbf{Predictors} & \textbf{Bootstrap frequency} \\
\hline
\multicolumn{1}{@{}l}{All-pairs}\\
\quad \texttt{PTPN1x9INV:Cyp11B2-5paINV } &0.83\\
\quad \texttt{Cyp11B2-5paINV:PTPN1i4 } &0.46\\
\quad \texttt{CD36x2aINV:MLRi6INV } &0.45\\
\quad \texttt{PTPN1i4INV:Cyp11B2-5paINV} & 0.40\\
\quad \texttt{Cyp11B2i4INV:PTPN1x9INV } &0.33\\
\quad \texttt{PTPN1x9INV:MLRI2V }&0.28\\
\quad \texttt{CD36x2aINV:KLKQ3E }& 0.27\\
\quad \texttt{Reached menopause?:AGT2R1A1166C} &0.22\\
\quad \texttt{MLRi6INV:Cyp11B2-5paINV } & 0.21\\
\quad \texttt{insulin t=60:KLKQ3E } &0.20
\\[3pt]
\multicolumn{1}{@{}l}{Screen: weak}\\
\quad\texttt{insulin t=-10:HUT2SNP7} & 0.96 \\
\quad\texttt{Reached menopause?:AGT2R1A1166C} & 0.96 \\
\quad\texttt{insulin t=-10:Cyp11B2i4INV} & 0.92 \\
\quad\texttt{insulin t=60:Cyp11B2i4INV} & 0.92 \\
\quad\texttt{CD36x2aINV:KLKQ3E} & 0.92 \\
\quad\texttt{Reached menopause?:insulin t=120} & 0.88 \\
\quad\texttt{insulin t=60:insulin t=120} & 0.88 \\
\quad\texttt{insulin t=-10:ADRB3W1R} & 0.88 \\
\quad\texttt{insulin t=-10:Cyp11B2-5paINV} & 0.88 \\
\quad\texttt{insulin t=120:ADRB3W1R} & 0.84 \\
\\[3pt]
\multicolumn{1}{@{}l}{Convex hierarchical test}\\
\quad\texttt{PTPN1x9INV:Cyp11B2-5paINV} & 0.63 \\
\quad\texttt{Reached menopause?:AGT2R1A1166C} & 0.47 \\
\quad\texttt{insulin t=60:KLKQ3E} & 0.34 \\
\quad\texttt{insulin t=-10:HUT2SNP7} & 0.33 \\
\quad\texttt{insulin t=-10:ADRB3W1R} & 0.32 \\
\quad\texttt{CD36x2aINV:MLRi6INV} & 0.31 \\
\quad\texttt{insulin t=60:Cyp11B2i4INV} & 0.26 \\
\quad\texttt{insulin t=-10:Cyp11B2i4INV} & 0.25 \\
\quad\texttt{CD36x2aINV:KLKQ3E} & 0.25 \\
\quad\texttt{Cyp11B2-5paINV:PTPN1i4} & 0.24 \\
\hline
\end{tabular*}
\end{table}

In Table~\ref{tab:sapphire3}, we present a bootstrap analysis to shed
light on the behavior of three methods: the all-pairs method,
the weak screening method considered in the simulation section,
and CHT. We record the top ten interactions appearing in the analysis
from each of 100 bootstrap samples.
The ten most frequently occurring interactions for each method are
shown in
Table~\ref{tab:sapphire3}. We see that there is one gene--gene
interaction that stands out for all-pairs, which includes it $83\%$ of
the time; interestingly, this interaction does not even appear in the
weak screening
method's list. The weak screening method cannot detect this
interaction because neither of the genes involved has a large enough main
effect. By contrast, in CHT this interaction is the most frequently
occurring of the interactions. This demonstrates CHT's greater
malleability with the hierarchy requirement: Large interactions can be
detected even if they have small main effect contrasts. This same
observation is true of the top three interactions in the all-pairs
list. 
Six interactions are shared between all-pairs and CHT;
all the interactions
appearing in
the CHT list but not in the all-pairs list involve clinical
variables (and are in the weak-screen list).

Finally, we note that only one of the top interactions found by our procedure were not
found in \citet{Park01012008}. However, this may not be surprising,
as their paper focused on multivariate modeling and conditional effects.

\section{Estimation of the false discovery rate}
\label{sec:fdr}
Permutations provide a convenient and robust way to estimate false
discovery rates in large-scale hypothesis testing. For example,
\citet{ST2012} devise a permutation scheme for the all-pairs
interaction test. In this scheme, one randomly assigns a component
of the interaction
contrast to group 1 or group 2 by flipping the sign of the
component at random.

This scheme can be easily adapted to the present setting:
The idea is to retain the main effect contrasts $w_j$ from the original fit
and to create randomized versions of the interactions. In particular,
let $f\dvtx \mathbb{R}^n\to\mathbb{R}^{p^2-p}$ represent the function of
the class
labels such that $z_{jk}=f(y)_{jk}$.\vspace*{1pt} For $b=1,\ldots,B$, we get
random permutations $y^{*(b)}\in\mathbb{R}^n$ of $y$ and compute
$z^{*(b)}=f(y^{*(b)})$.

Using \eqref{eq:teststat}, we get $\hat\lambda^{\prime *(b)}_{jk}$ based
on $(w, z^{*(b)})$.
Finally, we estimate the FDR as
%
\begin{equation}
\label{eq:fdr-hat}
\widehat{\mathrm{FDR}}(\lambda)= \min \biggl\{\frac{({1}/{B})\sum_{j,k,b}
I(\hat\lambda^{\prime *(b)}_{jk}>\lambda)}{\sum_{jk} I(\hat\lambda
'_{jk}>\lambda)},1 \biggr\}.
\end{equation}
Note that this estimate of FDR pools the null distributions from all
$jk$ pairs.
This kind of pooled null distribution is
commonly used, for example, in the SAM procedure [\citet{TTC01}]
and in  the aforementioned interaction test of \citet{ST2012}.
Its accuracy is quite high in simulation studies, although we know of
no rigorous results on its asymptotic properties.

Figure~6
shows the estimated FDR from this method for three of the scenarios
described earlier. We observe that the estimate is fairly accurate,
especially when the number of interactions called is small, but tends
to overestimate the true FDR by a moderate amount for larger numbers
of interactions called.
This may be due to the interdependence
of the test statistics $\hat\lambda_{jk}$ for each $j$.
Overestimation of the FDR corresponds to being conservative, which
is of less concern
than underestimation.

\begin{figure}

\includegraphics{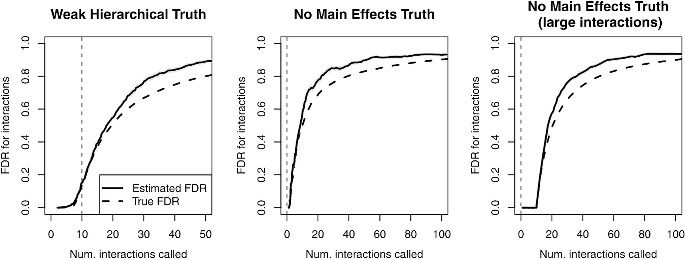}

\caption{Estimation of FDR for convex hierarchical testing
using permutations.
Result is an average over 50 simulations (with one standard error bars
shown as well). Vertical line is drawn at true number of nonzero interactions.}
\label{fig:fdrest0}
\end{figure}

In future work, it would be important to study the theoretical
properties of this permutation estimate.

\section{Discussion}
\label{sec:discussion}
We have proposed a hierarchical method for large-scale interaction
testing that biases its search toward interactions exhibiting at
least one moderate main effect. Our testing procedure is defined in
terms of a convex optimization problem but can be expressed in
closed form. Examination of the form of the statistic shows that it
incorporates hierarchy in a gentler way
than two-step procedures that screen out interactions based on main
effects. This distinction allows it to include large interactions
even when hierarchy is violated (as seen in the simulation).

This work could be generalized in several ways. We have focused
exclusively on
pairwise interactions: Extensions to $k$-way interactions, for $k>2$,
would bound the sum of such interactions by the size of the $k-1$ order
effect. With appropriate definitions for the interaction components,
$z_{jk}$, one could also
apply this procedure to interaction testing for the proportional
hazards model in survival analysis. More generally, the idea of
formulating a test statistic based on the knots of a convex
optimization problem's solution path may be of interest in contexts
beyond testing interactions.

\section*{Acknowledgments}
The authors would
like to thank the referees and two editors for comments that led to
improvements to this work.

\begin{supplement}[id=suppA]
\stitle{Supplement to ``Convex hierarchical testing of interactions''\\}
\slink[doi]{10.1214/14-AOAS758SUPP}
\sdatatype{.pdf}
\sfilename{aoas758\_supp.pdf}
\sdescription{We provide a detailed look at the optimization problem~\eqref{eq:hier} and prove all results in the paper.}
\end{supplement}

%



\printaddresses
\end{document}